\documentclass[structabstract]{aa}  
\usepackage{graphicx}
\usepackage{enumerate}
\usepackage{natbib}
\bibpunct{(}{)}{;}{a}{}{,} 
\usepackage{txfonts}
\usepackage[utf8]{inputenc}
\begin{document}
   \title{Inclinations of small quiet-Sun magnetic features based on a new geometric approach}

   \author{S.~Jafarzadeh\inst{1,}\thanks{Now at Institute of Theoretical Astrophysics, University of Oslo, Norway.}, S.~K.~Solanki\inst{1,2}, A.~Lagg\inst{1}, L.~R.~Bellot~Rubio\inst{3}, M.~van~Noort\inst{1}, A.~Feller\inst{1}, \and S.~Danilovic\inst{1}
             }

   \institute{Max Planck Institute for Solar System Research, Justus-von-Liebig-Weg 3, 37077 G\"{o}ttingen, Germany\\
   		\email{shahin.jafarzadeh@astro.uio.no}
         \and
             School of Space Research, Kyung Hee University, Yongin, Gyeonggi 446-701, Republic of Korea
         \and
          	 Instituto de Astrof\'{i}sica de Andaluc\'{i}a (CSIC), Glorieta de la Astronomia, s/n 18008 Granada, Spain\\             
             }

   \date{Received 14 January 2014 / Accepted 05 August 2014}

  \abstract
{High levels of horizontal magnetic flux have been reported in the quiet-Sun internetwork, often based on Stokes profile inversions.}
{Here we introduce a new method for deducing the inclination of magnetic elements and use it to test magnetic field inclinations from inversions.}
{We determine accurate positions of a set of small, bright magnetic elements in high spatial resolution images sampling different photospheric heights obtained by the Sunrise balloon-borne solar observatory. Together with estimates of the formation heights of the employed spectral bands, these provide us with the inclinations of the magnetic features. We also compute the magnetic inclination angle of the same magnetic features from the inversion of simultaneously recorded Stokes parameters.}
{Our new, geometric method returns nearly vertical fields (average inclination of around $14^{\circ}$ with a relatively narrow distribution having a standard deviation of $6^{\circ}$). In strong contrast to this, the traditionally used inversions give almost horizontal fields (average inclination of $75\pm8^{\circ}$) for the same small magnetic features, whose linearly polarised Stokes profiles are adversely affected by noise. We show that for such magnetic features inversions overestimate the flux in horizontal magnetic fields by an order of magnitude.}
{The almost vertical field of bright magnetic features from our geometric method is clearly incompatible with the nearly horizontal magnetic fields obtained from the inversions. This indicates that the amount of magnetic flux in horizontal fields deduced from inversions is overestimated in the presence of weak Stokes signals, in particular if Stokes $Q$ and $U$ are close to or under the noise level. Inversions should be used with great caution when applied to data with no clear Stokes $Q$ and no $U$ signal. By combining the proposed method with inversions we are not just improving the inclination, but also the field strength. This technique allows us to analyse features that are not reliably treated by inversions, thus greatly extending our capability to study the complete magnetic field of the quiet Sun.
}


  \authorrunning{S.~Jafarzadeh~et~al.}
  \titlerunning{Inclinations of small quiet-Sun magnetic features}
   \maketitle
%

\section{Introduction}
\label{sec-introincl}

The magnetic field strength distribution in the solar photosphere depends on the location: (1) in active and network regions $\mathrm{kG}$ fields dominate~\citep[e.g.][]{Stenflo1973,Solanki1993,Martinez-Pillet1997,Martinez-Gonzalez2012}, while (2) weak magnetic fields, up to a few $\mathrm{hG}$, are found all over the solar surface. The latter (i.e. weaker) component has been most intensely studied in internetwork (IN) areas, i.e. the supergranular interiors~\citep[e.g.][]{Livingston1971,Livingston1975,Lin1995,Khomenko2003,TrujilloBueno2004,Khomenko2005,Ishikawa2009,Beck2009,Ishikawa2011,BellotRubio2012}. Recently, however, $\mathrm{kG}$ fields have been found also in the IN~\citep{Lagg2010}. Compared to the active and network regions, the IN covers a much larger fraction of the solar surface and hence may contain most of the unsigned magnetic flux on the solar surface at any given time~\citep{SanchezAlmeida2004}.
Therefore, measuring reliable magnetic field properties of the IN areas is important for our understanding of, e.g. a local dynamo~\citep{Voegler2007,Danilovic2010a,Stenflo2012} and the dynamic coupling of the photosphere to the higher atmospheric layers~\citep{deWijn2009}. For recent reviews on the quiet-Sun IN magnetic fields, we refer to~\citet{deWijn2009}, \citet{Solanki2009}, and \citet{SanchezAlmeida2011}.

The characteristics of magnetic fields are traditionally inferred from the influence of the Zeeman effect on spectral lines. Observed Stokes profiles are treated through inversions of the polarised radiative transfer equations to determine, e.g. the strength and inclination angle of the magnetic field relative to the line of sight. Such a treatment depends, to a certain extent, on the model and sometimes on the initial parameters assumed in the employed inversion code. The inversion may fail (i.e. provide wrong results) when the polarisation signals are below or close to the noise level.

\begin{figure*}[th!]
\centering{\includegraphics[width=0.99\textwidth]{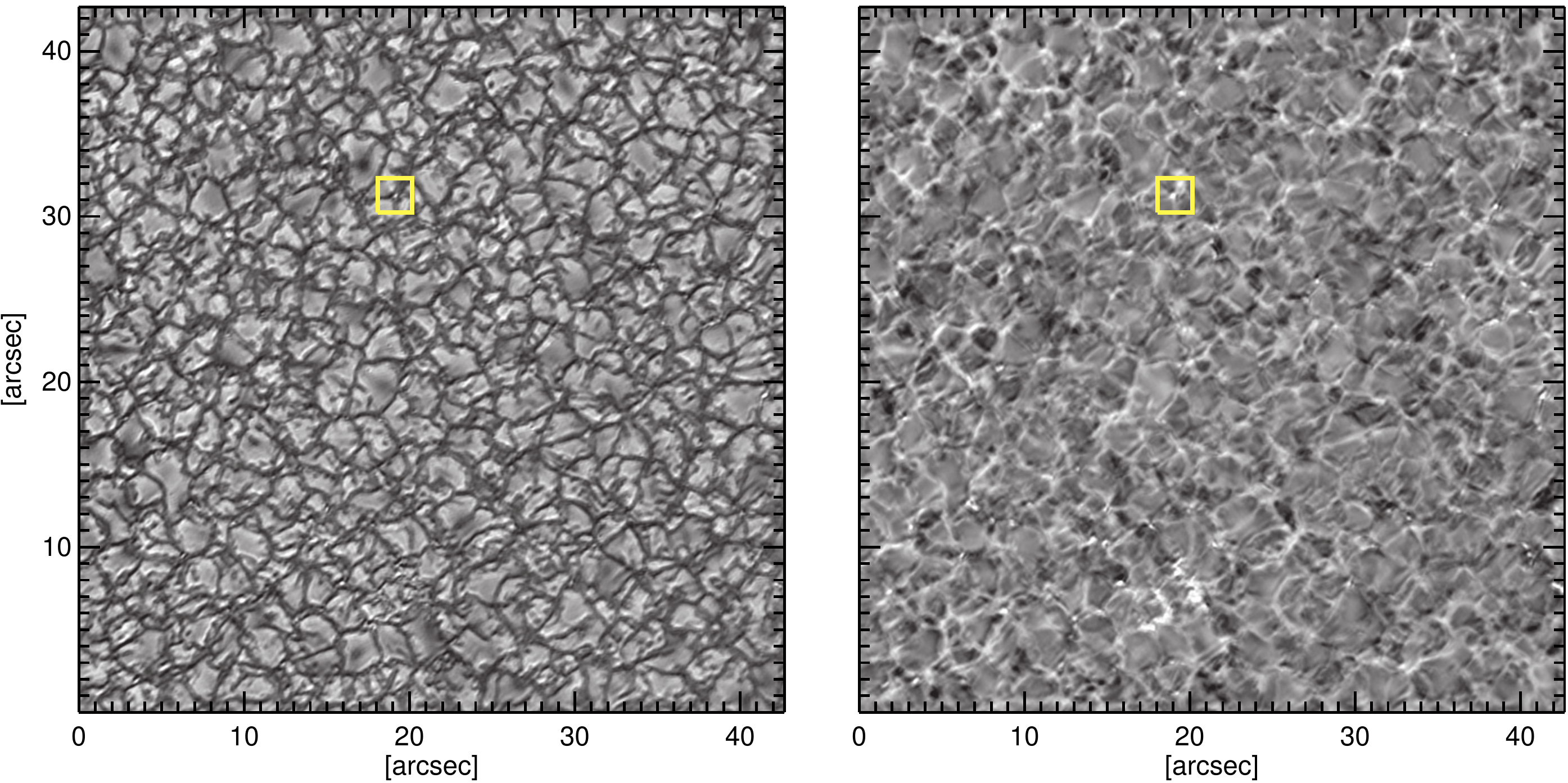}}
\caption{Example of co-spatial and co-temporal images from {\sc Sunrise}/IMaX.
Left: Full field of view (after apodization) of the IMaX at the continuum wavelength position next to the Fe~{\sc i} $5250.2~\AA$ line.
Right: Average of two line-positions at $-40~m\AA$ and $+40~m\AA$ from the Fe~{\sc i} $5250.2~\AA$ line-core.
The yellow box encloses a sample magnetic bright point (MBP) studied here.}
\label{fig:incl-obs}
\end{figure*}

High spatial resolution data have indicated that the magnetic field in IN areas is mainly horizontal~\citep{Lites1996,Orozco2007,Lites2008,Ishikawa2009,Orozco2012b}, while isotropic distributions of the magnetic field vector have also been reported~\citep{Asensio2009,Bommier2009}, and \citet{Stenflo2010,Stenflo2013} has argued that the field is on average more vertical than horizontal. It has been suggested that the origins of horizontal and vertical magnetic fields in the quiet Sun lie in the emergence of small bipolar loops and the concentration into flux tubes by convective collapse of the field forming their foot-points, respectively~\citep{Ishikawa2011}. A common origin of vertical and horizontal fields in the IN is predicted by modes of the local dynamo~\citep{Voegler2007,Schuessler2008}.

\citet{Borrero2011a,Borrero2012} have argued that excessive horizontal fields can be returned by inversions of noisy polarisation signals. They found horizontal and $\mathrm{hG}$ fields from inverting the observed Stokes profiles (in quiet-Sun IN regions), whereas their Monte-Carlo and numerical simulations revealed purely vertical fields with much weaker field strengths ($B<20~\mathrm{G}$) compared to those obtained from inversions. They found that only $30\%$ of the observed regions have sufficiently strong Stokes $Q$ and $U$ signals (i.e. signals sufficiently above the noise level) to allow the reliable determination of the magnetic field vector. \citet{deWijn2009}, \citet{Asensio2009} and \citet{Bommier2009} have all pointed out the difficulties in obtaining the magnetic inclination when the polarisation signal is weak and in particular, when the Stokes $Q$ and $U$ signals are not well above the noise level.

In contrast,~\citet{Orozco2012a},~\citet{Orozco2012b} and \citet{BellotRubio2012} argued that the mainly horizontal field in the IN deduced from inversions is consistent with their more up-to-date analyses or the higher signal-to-noise ratio data they analysed. This controversy highlights the need for an alternative method to determine the inclination of the magnetic vector independently of the strength of the Stokes $Q$ and $U$ profiles.

In this paper, we present a new and simple approach to obtain inclinations of small-scale bright magnetic structures (magnetic bright points; MBPs) using intensity images observed in different layers of the solar atmosphere. Since MBPs are co-located with the magnetic field, the inclination obtained in this manner should correspond to the inclination of the magnetic vector. We therefore call it the magnetic inclination proxy, $\gamma_{p}$. We test this correspondence by determining the inclination using the same method, but from the Stokes $V$ profiles sampling different heights (i.e. by comparing wavelengths at different distances from the line-core in a sufficiently strong spectral line). We use Stokes $V$ because of its larger signal-to-noise ratio (S/N) compared to other Stokes profiles. Furthermore, we compare the inclination angles obtained from our proposed geometric method and those computed from Stokes inversion codes.

The plan of this paper is as follows: in Sect.~2 we briefly review the data used for the study. In Sect.~3 we describe the technique used to measure the magnetic field inclination along with the results. We also introduce the inversion codes that we apply to the polarimetric data and discuss the comparison between differently measured inclination angles. We summarise our conclusions in Sect.~4.

\section{Observational data}
\label{sec-obsincl}

For most of this study we use high spatial resolution observations acquired on 9 June 2009 (between 01:32:06 and 01:58:43 UT), in the quiet-Sun disc-centre, by the Imaging Magnetograph eXperiment (IMaX;~\citealt{Martinez-Pillet2011}) on board the {\sc Sunrise} balloon-borne solar observatory~\citep{Solanki2010,Barthol2011,Berkefeld2011}.

The data consist of the full Stokes vector ($I$,$Q$,$U$~and~$V$) measured in five wavelength positions located at $-80$, $-40$, $+40$, $+80~\mathrm{m\AA} $ and $+227~\mathrm{m\AA} $ from the centre of the magnetically sensitive Fe~{\sc i}~$5250.2~\mathrm{\AA}$ line. The first four wavelengths lie within the line, while the last one samples the continuum. The image sequences were obtained at a cadence of $33~\mathrm{sec}$ with a noise level of $\approx3\times10^{-3}~I_{c}$ after phase-diversity (PD) reconstruction ($V5-6$ level~$2$ data;~\citealt{Martinez-Pillet2011}). We also analysed data prior to PD reconstruction (level~$1$ data), which were flat-fielded and corrected for instrumental effects. They have a noise level of $\approx10^{-3}~I_{c}$. The $1\sigma$ noise levels of all Stokes profiles are summarised in Table~\ref{table:incl-noise} (discussed in detail below).

We prepared two datasets of intensity images, each sampling a different height above the solar surface. We used these two sets of images for determining the proxy of the magnetic field's inclination angle $\gamma_{p}$ from measurements at the two heights. One set is composed of IMaX Stokes $I$ continuum images, which sample the continuum formation height. To obtain the second set we form images corresponding to a combination of the line core and the line's inner flanks by averaging the $-40$ and $+40~\mathrm{m\AA} $ wavelength positions of the IMaX Stokes $I$ normalised to the continuum intensity. We refer to these images as line-core images in the following. Figure~\ref{fig:incl-obs} shows example frames of the IMaX continuum intensity (left) and the IMaX line-core (as described above; right).

\begin{figure}[h!]
	\centering
	\includegraphics[width=8.2cm, trim = 0 0 0 0, clip]{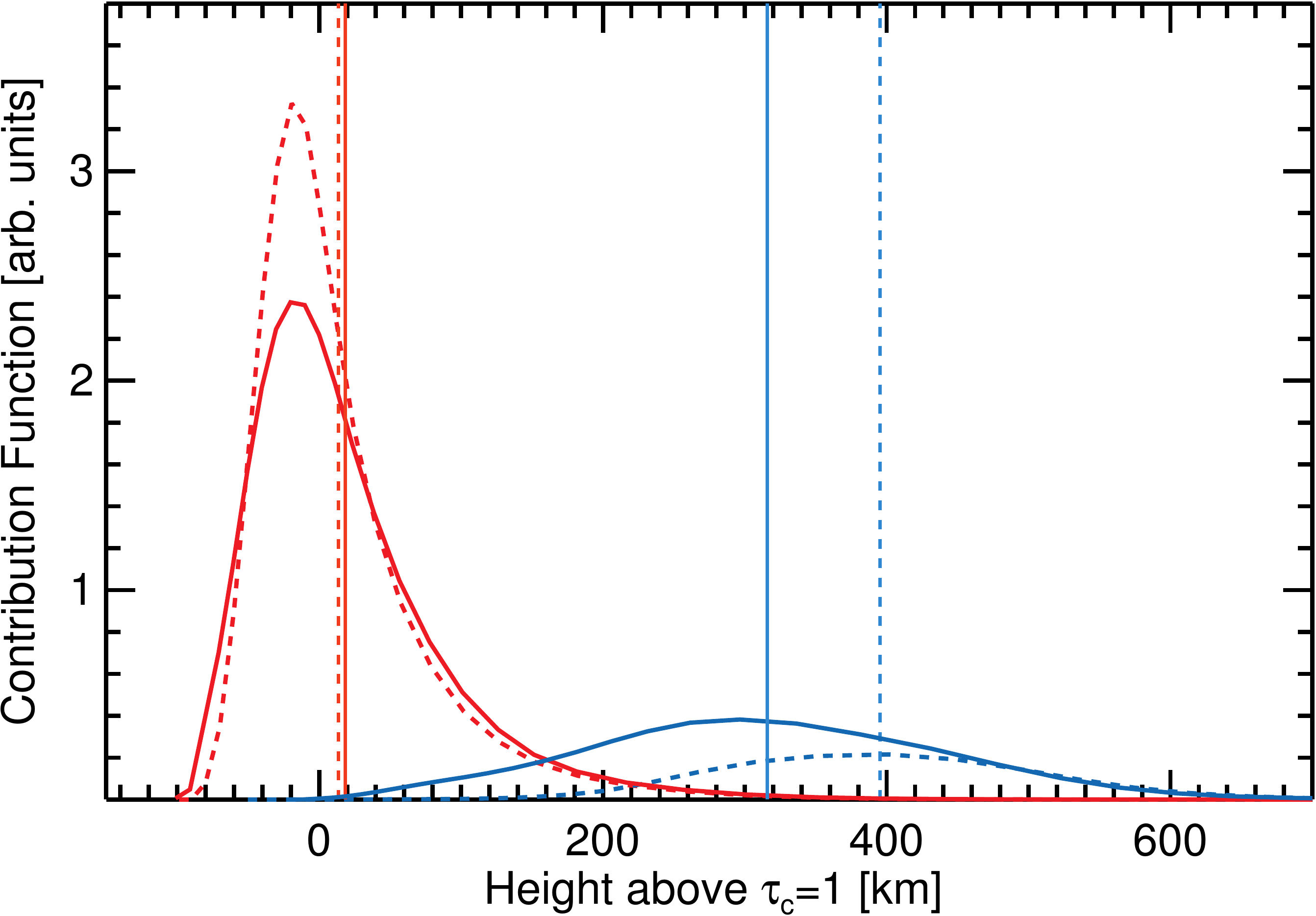}
	\caption{Line depression contribution functions for the {\sc Sunrise}/IMaX continuum (red) and line-core (blue) positions (see main text). The dashed and solid lines represent the FALC and FALP model atmospheres, respectively. The vertical lines indicate the corresponding weighted mean formation heights (heights above continuum optical depth unity, $\tau_{c}$).
}
	\label{fig:incl-fh}
\end{figure}

We used the RH radiative transfer code of \citet{Uitenbroek2001} to estimate the formation heights of the two layers sampled by the data products introduced in the last paragraph. We employed the code in the LTE mode for two atmospheric models, FALC and FALP, representing the averaged quiet-Sun and plage regions (or magnetic bright points; MBPs), respectively~\citep{Fontenla1993,Fontenla2006}. Plotted in Fig.~\ref{fig:incl-fh} are the line depression contribution functions (CFs; \citealt{Magain1986,GrossmannDoerth1988}) versus height above continuum optical depth unity ($\tau_{c}=1$; optical depth at $5000~\mathrm{\AA} $), obtained by integrating the computed, IMaX filter-profile weighted, CFs over wavelength, for IMaX continuum (red) and IMaX line-core (blue) and for FALC (dashed line) and FALP (solid line) model atmospheres. The vertical lines indicate the mean formation heights. These formation heights result in a $\sim300^{+100}_{-200}~\mathrm{km}$ height difference between the two layers in magnetic elements (see Sect.~\ref{subsec-incl-uncertainty} for an estimate of the uncertainty).

\noindent
For a part of the analysis, in order to increase the S/N of the often weak polarisation signals found in the quiet-Sun, we average the Stokes $V$ signals over the four wavelength positions inside the line, normalised to the local quiet-Sun continuum intensity ($I_{c}$). To avoid cancellation, the sign of the two red wavelength points are reversed prior to averaging. We refer to this quantity as $CP$ in this paper. Furthermore, we form the total linear polarisation ($LP$) from the Stokes $Q$ and $U$ signals (i.e. $\sqrt{Q^2+U^2}$). Similar to $CP$, we aim to increase the S/N of the $LP$ by averaging over the four wavelength positions. However, because squaring the Stokes profiles also squares the noise, which then no longer cancels through averaging, we do not form the $LP$ for each single wavelength position prior to averaging. The $LP$ is instead computed as
\begin{equation}
LP = \frac{1}{2\: I_{c}}\left ( \sqrt{\overline{Q}_{14}^{2} +\overline{U}_{14}^{2}} + \sqrt{\overline{Q}_{23}^{2} + \overline{U}_{23}^{2}} \right )\,,
	\label{equ:LP}
\end{equation}
\noindent
where $\overline{Q}$ and $\overline{U}$ are respectively the averaged Stokes $Q$ and $U$ profiles over two wavelength positions with the same sign under normal circumstances, i.e. the outermost (indicated by indices $1$ and $4$) and the innermost (represented by indices $2$ and $3$) wavelength positions in the line. The outer and inner wavelength positions are treated separately in order to avoid possible cancellation.

The $CP$ and $LP$ are measures of the level of polarisation and are only used as selection criteria.\\
In addition to using the PD reconstructed data, we also determined the $CP$ and the $LP$ from the less noisy non-reconstructed data (but at the cost of a factor of $2$ lower spatial resolution compared to the PD reconstructed data).

An average $1\sigma$ noise level of $\approx6\times10^{-4}$ was obtained for $CP$ and $LP$ from the non-reconstructed data. The noise levels were determined as the standard deviations at the continuum position since we do not expect any polarisation signal in the continuum. The continuum noise level was taken to be at each wavelength point. By combining wavelength positions, following above determinations of $CP$ and $LP$, the reduced noise level is obtained. The Stokes $V$ signals found by~\citet{Borrero2010} at this wavelength are restricted to sufficiently few spatial locations not to influence the noise estimate.

We also spatially smooth the $CP$ and $LP$ maps obtained from non-reconstructed data by applying a boxcar average of $3\times3$ pixels, to additionally reduce the noise in the weakly polarised regions under study without degrading the spatial resolution (the non-reconstructed data are oversampled by this amount). After this step, the $1\sigma$ noise levels for both $CP$ and $LP$ are reduced to $\approx3\times10^{-4}$ in the non-reconstructed maps. A spatial smoothing by $9$ pixels would be expected to reduce the noise level by a factor of $3$ under ideal conditions. However, we found the actual reduction in the noise level is smaller due to the presence of coherent noise caused by image jitter.

In the following, we will use the $CP$ and $LP$ maps only to extract and distinguish the polarisation signal from the noise. We note here that the individual profiles to be inverted have worse noise levels than the integrated $CP$ and $LP$ maps. By retrieving the magnetic field vector through inversion of Stokes profiles with different noise levels, we will be able to investigate the effect of noise level on the inclination of the magnetic vector, although it must be borne in mind that the data with different noise levels do differ in spatial resolution and sampling from each other.

To summarise, the differently treated datasets are: ($1$) The non-reconstructed (but flat-fielded and corrected for instrumental effects) images, ($2$) spatially smoothed non-reconstructed data, obtained by applying a $3$-pixel boxcar smoothing, and ($3$) the higher spatial resolution (by a factor of $2$) PD reconstructed images with a higher noise level than the non-reconstructed data. Table~\ref{table:incl-noise} summarises the $1\sigma$ noise levels of $CP$, $LP$ as well as individual of Stokes profiles for these differently treated datasets.

\begin{table}[hb]
\caption{Summary of $1\sigma$ noise levels in the employed data sets}     
\label{table:incl-noise}      
\centering                   
\begin{tabular}{l c c c}      
\hline\hline \\  [-1.9ex]        
Parameter & PDR\tablefootmark{$*$} & NR\tablefootmark{$*$} & SSNR\tablefootmark{$*$}\\
\\ [-1.9ex]
\hline \\ [-2.0ex]              
   $Q/I_{c}$ & $2.6\times10^{-3}$ & $8.3\times10^{-4}$ & $4.6\times10^{-4}$\\   
   $U/I_{c}$ & $3.6\times10^{-3}$ & $1.1\times10^{-3}$ & $4.8\times10^{-4}$\\
   $V/I_{c}$ & $3.3\times10^{-3}$ & $1.0\times10^{-3}$ & $6.3\times10^{-4}$\\
   $CP$ & $1.7\times10^{-3}$ & $5.2\times10^{-4}$ & $3.2\times10^{-4}$\\
   $LP$ & $2.2\times10^{-3}$ & $6.6\times10^{-4}$ & $2.9\times10^{-4}$\\
\\ [-2.0ex]
\hline
\end{tabular}
\tablefoot{
\tablefoottext{$*$}{PDR: Phase diversity reconstructed data; NR: Non-reconstructed data; SSNR: Spatially smoothed Non-reconstructed data.}
}
\end{table}

\section{Analysis and results}

In this section we introduce and describe a simple method for determining the magnetic field's inclination angle proxy from high resolution intensity images which works for magnetic features that produce brightness enhancements. We manually select these bright elements, whose $CP\geq4\sigma$, in intensity images of differently treated data.
In parallel to our measurements at two heights, we also determine the magnetic field vector in the selected features by inverting the observed Stokes profiles.

\subsection{Inclination from measurements at two heights}
\label{subsec-geometricmethod}

The new approach is based on the fact that in addition to the clearly present low-lying loops~\citep[e.g.][]{Martinez-Gonzalez2007,Ishikawa2008,Ishikawa2009,Martinez-Gonzalez2009,Danilovic2010b} and a probably truly turbulent field~\citep[e.g.][]{PietarilaGraham2009} there are also magnetic fields that are better represented by slender flux tubes~\citep[e.g.][]{Solanki1996b}. Flux tubes describe relatively isolated, concentrated magnetic fields, so-called magnetic elements (MEs). They are rooted in the solar interior and unless they have freshly emerged, they extend into the upper atmospheric layers~\citep{Stenflo1989,Solanki1993}. The cross-section of an intense and thin ME in each layer of the photosphere manifests itself as a bright point (MBP) in intensity images, due to a combination of continuum enhancement and line weakening~\citep[e.g.][]{Keller1992,Kiselman2001,Nagata2008}. In the lower photosphere the excess brightness of MBPs is due to radiation from subsurface hot walls of the flux tubes (\citealt{Spruit1976}; cf. e.g.~\citealt{Deinzer1984b}), whereas in the middle photosphere and higher layers it is produced by radiative and non-radiative heating. We take the centre of gravity of the intensity patches of the observed MBPs to represent the locations of the centres of MEs.

Therefore, connecting the MBPs identified in well-aligned intensity images corresponding to two different atmospheric layers (e.g. the solar surface and an upper photospheric layer) can provide its inclination, if they belong to the same ME and if the height difference between the two layers is known. We refer to this inclination angle as $\gamma_{p}$ ($p$ for proxy) to distinguish it from the true magnetic inclination $\gamma$.
We apply this method to small MBPs identified in {\sc Sunrise}/IMaX high spatial resolution images in the Fe~{\sc i}~$5250.2~\mathrm{\AA} $ continuum and in the line-core (as described in Sect.~\ref{sec-obsincl}).

\begin{figure}[h!]
	\centering
	\includegraphics[width=9cm]{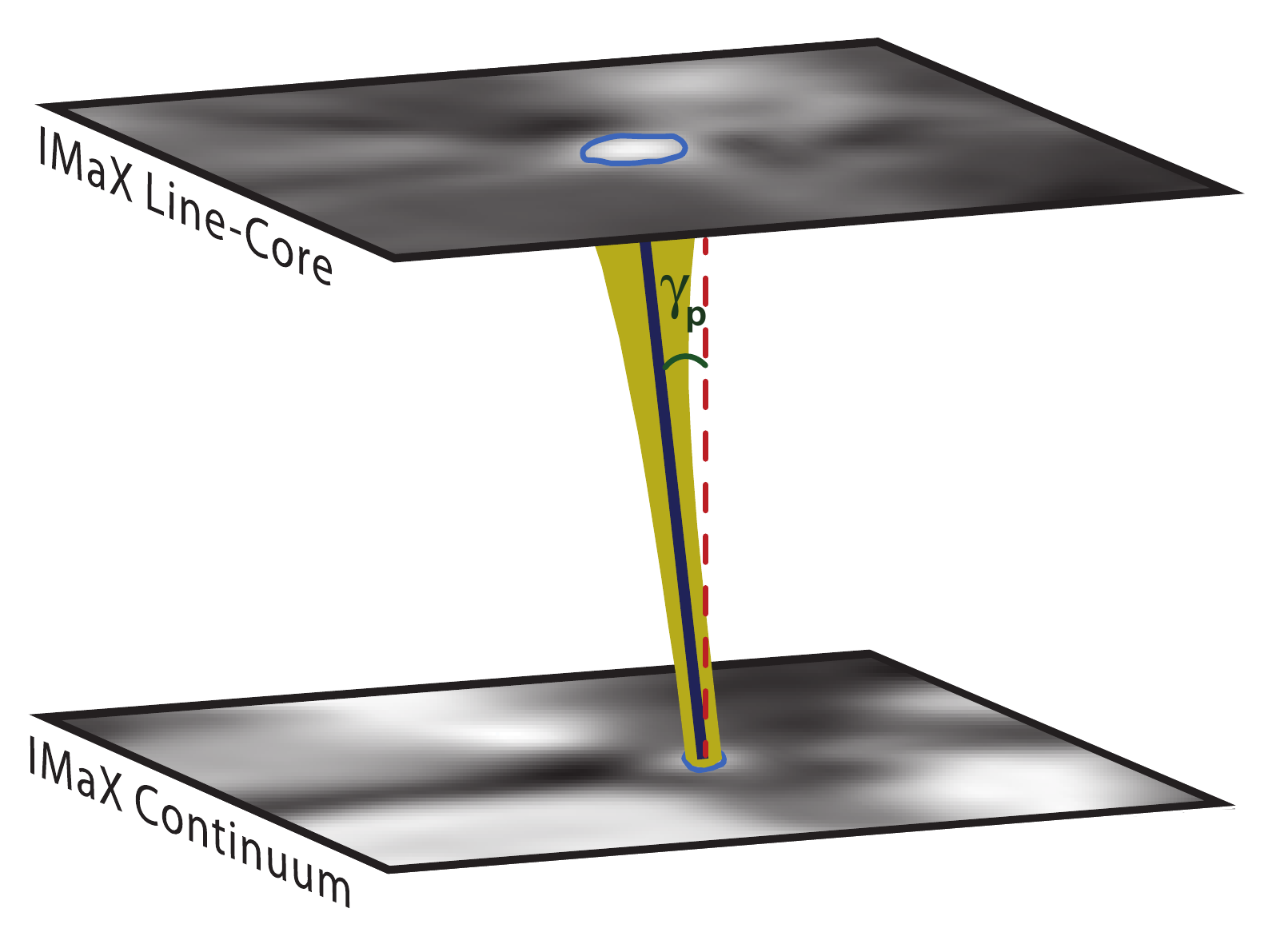}
	\caption{A magnetic bright point (MBP; marked in Fig.~\ref{fig:incl-obs}) in the continuum and line-core intensity images obtained by {\sc Sunrise}/IMaX. A schematic flux tube connecting the locations of the MBP at the two photospheric layers (yellow shading) with an approximate height difference of $300^{+100}_{-200}~\mathrm{km}$ has been added. The inclination of the axis of this flux tube (solid line) relative to the surface normal (dashed line) is marked as $\gamma_{p}$. The figure is a simplification in that the two images are assigned fixed heights, with the vertical axis expanded by a factor of $2.8$. The blue contours outline the MBP at the two atmospheric layers.}
	\label{fig:inclBP1}
\end{figure}

The isolated, small, and magnetic bright points are manually selected in both continuum and line-core images. In IMaX line-core images they are required to meet the criteria described in ~\citet{Jafarzadeh2013a} (hereafter referred to as Paper~I) which they applied to Ca~{\sc ii}~H images. The application of the same criteria is reasonable for IMaX line-core filtergrams whose intensity variations are similar to the Ca~{\sc ii}~H images analysed in Paper~I. We checked this by applying these criteria to overlapping SuFI~\citep{Gandorfer2011} Ca~{\sc ii}~H images and IMaX line-core images, obtaining a good match. In addition, a MBP is considered magnetic if it is located inside a magnetic patch (i.e. with $CP\geq4\sigma$ noise level). Therefore, the selected MBPs (a total of $386$) only include very small magnetic, bright, point-like features in quiet-Sun IN areas, which facilitates an accurate locating procedure. Many of these small MBPs are likely spatially unresolved by {\sc Sunrise} \citep{Jafarzadeh2013a,Riethmuller2013b}. Larger MBPs that are also found in {\sc Sunrise}/IMaX \citep{Lagg2010,Riethmuller2010} are not considered in this study, since their (normally) non-uniform brightness and internal fine-structure make finding a unique and accurate position less straightforward. The precise location of the MBP is computed using an algorithm described in detail in Paper~I. An important difference here, compared to~\citet{Jafarzadeh2013a}, is that the algorithm crops a small area (i.e. a square with sides of $0.6~\mathrm{arcsec}$) from each frame containing the manually selected MBP roughly at the centre, prior to the locating process. This facilitates the procedure, particularly in continuum images. This algorithm determines the location of the MBP, i.e. centre of gravity of intensity down to sub-pixel accuracy. Once the MBP has been located in the upper layer, we then search for a counterpart in the continuum images. We look for compact MBPs in intergranular lanes. The two closest MBPs at each height are assigned to each other, with the further requirement that the Stokes $V$ signal supports this identification.

Figure~\ref{fig:inclBP1} illustrates the identification at two heights of the MBP marked in Fig.~\ref{fig:incl-obs} and the interpolation of the flux tube between the two heights. The locations of the MBPs in the two layers, with a height difference of $\approx300^{+100}_{-200}~\mathrm{km}$, are connected by a solid line. The yellow shaded surface illustrates very schematically the flux tube's expansion based on the determined MBP areas in the continuum and the line-core images. The expansion factor of $\approx2.8$ is comparable to that obtained from the network flux-tube model of \citet{Solanki1986} and the thin-tube approximation \citep{Defouw1976}. We have used the same criteria to measure the size of the MBP in both atmospheric layers, so that the expansion factor should not be affected by that, as long as the features do not lie below the resolution limit (in which case the expansion factor would likely be underestimated). If the intensity contrast changes strongly with height, the determined expansion factor can also be affected.

The (height averaged) inclination of such a flux tube is then computed by simply dividing the offsets between the locations of the MBP in the continuum and the line-core images by the estimated height difference between those layers. The flux tube illustrated in Fig.~\ref{fig:inclBP1} has an inclination angle of $\gamma_{p}=19^{+11}_{-8}$~degrees (see Sect.~\ref{subsec-incl-uncertainty} for an estimate of the uncertainty).

\subsubsection{Sources of uncertainty}
\label{subsec-incl-uncertainty}

Uncertainty in the determined inclination angles is induced by biases from different sources:

\textit{Location of MBP}: The technique used to locate the MBPs has an accuracy of $0.05$ pixel at best (i.e. when the MBP is ideally small; see Paper~I for details). We employ a more conservative value of $0.5$ pixels for the uncertainty, which takes the effect of the MBPs' size and intensity variations into account~\citep{Jafarzadeh2013a}. The IMaX pixel size of $0.0523~\mathrm{arcsec}$ thus implies an uncertainty of $\sim19~\mathrm{km}$, in the horizontal plane.

\textit{Estimate of the height difference}: The mean formation heights of the IMaX continuum and line-centre (as described in Sect.~\ref{sec-obsincl}) depend on the atmospheric structure, which leads to an uncertainty in the difference between the heights of the two sampled layers ($\approx300~\mathrm{km}$). A value of $\pm100~\mathrm{km}$ for this uncertainty, which corresponds to a relative uncertainty of $33\%$, can be considered to be conservative, in the sense that we clearly observe reversed granulation in the upper atmospheric layer (see Fig.~\ref{fig:incl-obs}; right panel), which is only seen at heights greater than $\approx250~\mathrm{km}$ \citep{Wedemeyer2004}. Hence, we do not expect a height difference smaller than $200~\mathrm{km}$ between the two atmospheric layers over most of the field of view. However, within magnetic elements the Fe~{\sc i}~$5250.2~\mathrm{\AA}$ line is strongly weakened, owing to its low excitation potential. This leads also to a lower than average formation height within magnetic elements. With this consideration in mind we keep $+100~\mathrm{km}$ (formation height increase), but set $-200~\mathrm{km}$ (formation height decrease). This means that we allow for a formation of the line only $100~\mathrm{km}$ above the continuum formation level averaged over the magnetic element. We note that if the height difference would be larger than
$400~\mathrm{km}$, then our method would overestimate the inclination angles (i.e. the fields would in reality be more vertical than what our method returns).

\textit{Observing-time difference}: We formed the line-centre images by averaging the innermost wavelength positions of the IMaX Stokes $I$ normalised to the continuum point (Sect.~\ref{sec-obsincl}). The datasets we used here have a cadence of $33~\mathrm{sec}$, meaning an average time-difference of $\sim16.5~\mathrm{sec}$ between the IMaX line-centre and the IMaX continuum images. MBPs of the type under study move horizontally with an average speed of $2.2~\mathrm{km}\, \mathrm{s}^{-1}$~\citep{Jafarzadeh2013a}. Assuming that the MBP moves at this speed in a fixed direction during the difference in time between the images introduces on average an offset of $\approx36~\mathrm{km}$ between the locations of the MBPs at the two heights.

The combination of the three uncertainties mentioned above translates into an uncertainty in the inclination angle $\gamma$ determined from measurements at the two heights. A combination of $40~\mathrm{km}$ uncertainty in the horizontal plane ($\sigma_{h}$) as well as $-200/+100~\mathrm{km}$ uncertainties in measuring the height difference between the two layers ($\sigma_{z}$) leads to an average uncertainty of $\sigma_{\gamma}\approx-8^{\circ}/+11^{\circ}$ in the determined inclination angles ($\sigma_{\gamma}\approx\sigma_{h/z}/(1+h^2/z^2)$), where $h$ and $z$ are the mean values of the horizontal offset and the height difference between the two layers. $\sigma_{h/z}$ is the uncertainty of $h/z$ determined using $\sigma_{h}$ and $\sigma_{z}$.

Other sources of error are Doppler shifts and Zeeman splitting, which can cause the parts of the spectral lines in the IMaX filters to sample a different height than in the absence of these effects. In order to estimate the errors introduced by these effects in a statistical sense we also compared the Ca~{\sc ii}~H line-core sampled by a $1.8~\mathrm{\AA} $ broad SuFI filter~\citep{Gandorfer2011}, which is sufficiently broad not to be affected by typical Doppler shifts or Zeeman splitting. Typical formation heights were determined by~\citet{Jafarzadeh2013a} to be $\approx500~\mathrm{km}$ above $\tau_{c}=1$ based on computations of contribution functions. The height of formation estimated from phase difference of acoustic oscillations observed in the SuFI Ca~{\sc ii}~H channel and in the lower photosphere confirms this value (Jafarzadeh et al. in preparation). Only the subset of MBPs lying in the narrower SuFI field of view can be analysed in this way. Moreover, a careful sub-pixel alignment of the SuFI images to those from IMaX was carried out. The reversed granulation visible in both, the Fe~{\sc i} line-core images obtained by IMaX and the SuFI Ca~{\sc ii}~H images, allowed this alignment to be performed to better than a SuFI pixel. By comparing the centre of gravity of MBP locations in the SuFI Ca~{\sc ii}~H and in the IMaX continuum, we also obtain $\gamma_{p}$ values. The average and standard deviation of these $\gamma_{p}$ are $7^{\circ}$ and $4^{\circ}$, respectively. This agrees well with the values obtained from the IMaX line-core and continuum ($14^{\circ}$ and $6^{\circ}$, respectively; see Sect.~\ref{statisticsIncl}). The smaller inclination obtained from the SuFI/Ca~{\sc ii}~H-IMaX/continuum combination may have a variety of causes. Firstly, the height difference between Ca~{\sc ii}~H and the $5250.4~\mathrm{\AA} $ continuum is larger, reducing the effect of errors/uncertainties in deducing the horizontal position on the derived $\gamma_{p}$. Secondly, because of the higher cadence of SuFI the time difference between the images at the two heights is smaller, thus reducing uncertainties due to the motions of magnetic features. Finally, unlike the $5250.2~\mathrm{\AA} $ line-core intensity sampled by IMaX, the Ca~{\sc ii}~H intensity provided by SuFI is almost unaffected by (reasonable) Doppler shifts and Zeeman splitting, which could affect the location of the brightness peak.

Although the MBPs are manifestations of magnetic elements, the location of the centre of gravity of intensity may have an offset with respect to the strongest magnetic field. This can also bias the measured inclination angles in the magnetic elements based on the detected MBPs in intensity images. Since we studied small MBPs observed close to the quiet-Sun disc-centre we expect such an effect to be relatively small. We tested it by additionally considering Stokes $V$ images obtained at the four wavelength positions at which {\sc Sunrise}/IMaX measures within the spectral line. These wavelengths sample four atmospheric layers between the IMaX continuum and the line-centre, whereby they form two pairs of wavelengths with the two wavelengths in a pair having similar formation heights. An average height difference of $\approx43~\mathrm{km}$ between the two heights was estimated by computing the contribution functions of the FALP model atmosphere at the two IMaX wavelength positions $+40$ and $+80~\mathrm{m\AA} $ from the line-centre, after convolving the spectra with the transmission profile of the IMaX Fe~{\sc i} filter (using the same code as described in Sect.~\ref{sec-obsincl}).

This method is expected to be less certain, because of the smaller height difference available (for the IMaX data). Moreover, it cannot be completely ruled out that the distribution of field strengths changes with height, although small-scale magnetic features behave rather like the second-order thin-tube approximation\footnote[1]{The thin flux tube approximation is based on an expansion method about the tube axis. The second-order approximation considers the expansion up to the second order, which allows for a first-order radial magnetic and velocity field and azimuthal components of the velocity and magnetic fields~\citep{Ferriz-Mas1989b}.}, at least in 3D MHD simulations \citep{Yelles-Chaouche2009}, so that the location within the feature of the peak in the field strength does not change significantly with height. To find out how strongly such a change in location with height would affect our results we did a simple Monte Carlo test, by generating a set of simple $150~\mathrm{km}$ wide flux tubes with a normal distribution of inclination angles centred on $14^{\circ}$ with a standard deviation of $6^{\circ}$. Then, assuming that the maximum field strength/brightness lies at opposite walls of the flux tubes at the two heights separated by $300~\mathrm{km}$, we recomputed the distribution of inclination angles, and hence, their mean value. We finally applied our method to 10000 realisations. The thus obtained distribution of inclinations to the vertical has a mean value of $25^{\circ}$ (with a standard deviation of $12^{\circ}$), i.e. $11^{\circ}$, on average, larger than the original. Even the extreme change in the location of the peak of the brightness or magnetic field strength introduced in this test changes the returned average inclination by a small amount that is only slightly larger than the scatter of the inclination values (individual inclinations, however, are affected more strongly).

\begin{figure*}[th!]
\centering{\includegraphics[width=0.99\textwidth]{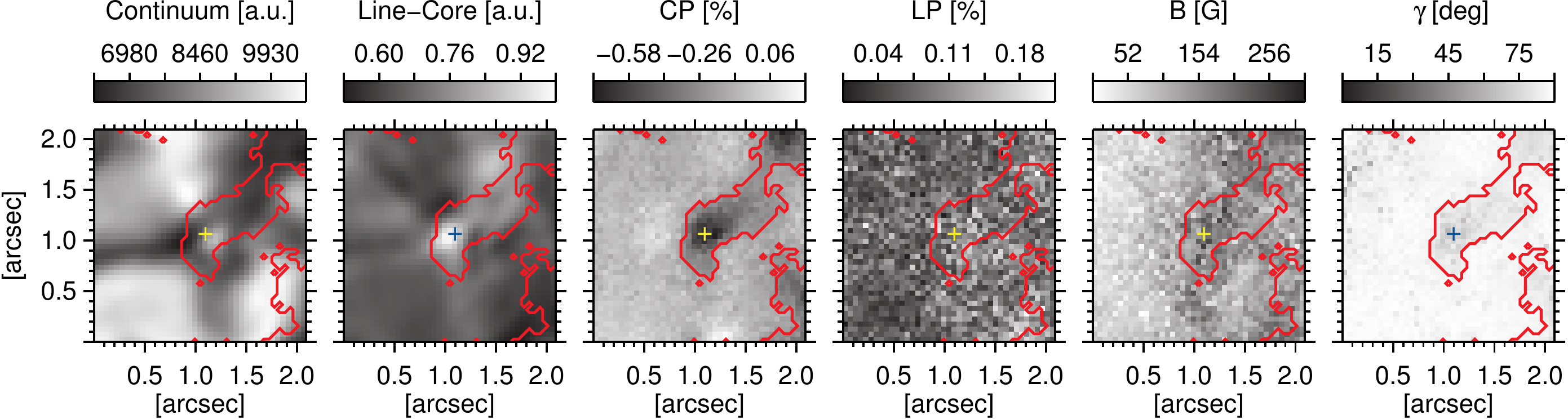}}
\caption{Example of a small magnetic bright point (MBP) marked in Fig.~\ref{fig:incl-obs}.
From left to right: IMaX continuum intensity, IMaX line-core (see main text for a description), a measure of the strength of Stokes $V$ ($CP$), total net linear polarisation ($LP$; see main text), magnetic field strength ($B$) and field inclination ($\gamma$). The magnetic field parameters $B$ and $\gamma$ are computed using the SPINOR inversion code. Red contours outline magnetic regions and match the $4\sigma$ noise level in $CP$. Location of the centre of gravity of the $CP$ distribution is indicated by a cross in all panels to facilitate comparing different maps.
}
\label{fig:subframes}
\end{figure*}

The centre of gravity of magnetic patches (i.e. Stokes $V$ signals greater than $4\sigma$ noise level) at the rough positions of the MBPs (detected in the intensity images) was considered as the location of magnetic elements. The peak Stokes $V$ signal in the magnetic element tends to have a spatial offset of $0.6~\mathrm{pixel}$ on average from the location of the centre of gravity of the Stokes $V$ patch. The centre of gravity of the magnetic patches has an offset of $1-1.5~\mathrm{pixels}$ on average from that determined for the MBPs in the IMaX Stokes $I$ images (both in the continuum and in the line-core). This implies an average offset of $48~\mathrm{km}$ between the magnetic and brightness structures, which is of the same order of the other uncertainties in the horizontal location described above. We note that the IMaX continuum samples a lower atmospheric layer compared to the four wavelength positions in the line. Since the magnetic field patches are normally bigger and more amorphous than the MBPs, their centres of gravity were measured within a small circle (diameter of $\approx0.4~\mathrm{arcsec}$) centred at the pixel with maximum value. The inclination was then determined from the location of the Stokes $V$ magnetic patches (sampled at the two heights; with an estimated height difference of $43~\mathrm{km}$) in a manner similar to our proposed approach for the intensity images, described earlier. The inclinations of the flux tubes determined by connecting the centre of gravity of same magnetic elements in the Stokes $V$ images made in the inner and outer flanks of the line ($22^{\circ}$ on average; with a distribution's standard deviation of $19^{\circ}$) are consistent with those obtained from IMaX intensity images sampling two heights ($14^{\circ}$ on average; see Sect.~\ref{statisticsIncl}). The values obtained from Stokes $V$ directly have a larger uncertainty due to the significantly smaller height difference.

\subsection{Inversions}
\label{sect:inversionmethod}

In order to retrieve the magnetic field strength ($B$) and the field inclination angle ($\gamma$) from the observed polarisation signals, we use the results of three inversion codes: ($a$) SPINOR~\citep{Frutiger2000,Berdyugina2003} which computes the synthetic Stokes profiles based on the STOPRO routines~\citep{Solanki1987}, ($b$) SIR~\citep{Ruiz-Cobo1992}, and ($c$) VFISV~\citep{Borrero2011b}. Both the SPINOR and SIR codes numerically solve the polarised radiative transfer equations under the assumption of local thermodynamic equilibrium (LTE) and iteratively minimise the difference between the computed and the observed profiles by modifying the initial model atmosphere using a response function-based Levenberg-Marquardt algorithm. 
The Harvard Smithsonian Reference Atmosphere (HSRA;~\citealt{Gingerich1971}) was used as the initial model atmosphere for both inversion codes. The temperature in the SPINOR code is modified with three nodes at $log \tau_{5000~\mathrm{\AA} }=0$, $-0.9$ and $-2.5$, while the SIR results are based on a temperature stratification in the range of $-4<log \tau_{5000~\mathrm{\AA}}<+1.4$ with two nodes in the temperature. The other parameters (i.e. $B$, $\gamma$, azimuth angle, line-of-sight velocity, and micro-turbulent velocity) are height independent in both SPINOR and SIR codes. For details on the SIR inversion carried out on the same data as used in this paper, we refer the reader to~\citet{Guglielmino2012}.
The VFISV code analytically solves the radiative transfer equation based on the Milne-Eddington approximation of the solar atmosphere. A set of initial parameters are iteratively modified by all codes until the best match between the synthetic and observed Stokes profiles is achieved. 
A magnetic filling factor of unity has been assumed for all inversions.

We will refer to the inclination angles computed by the inversions as $\gamma_{i}$ ($i$ for inversion).

\begin{figure}[th!]
    \centering
	\includegraphics[width=8.0cm]{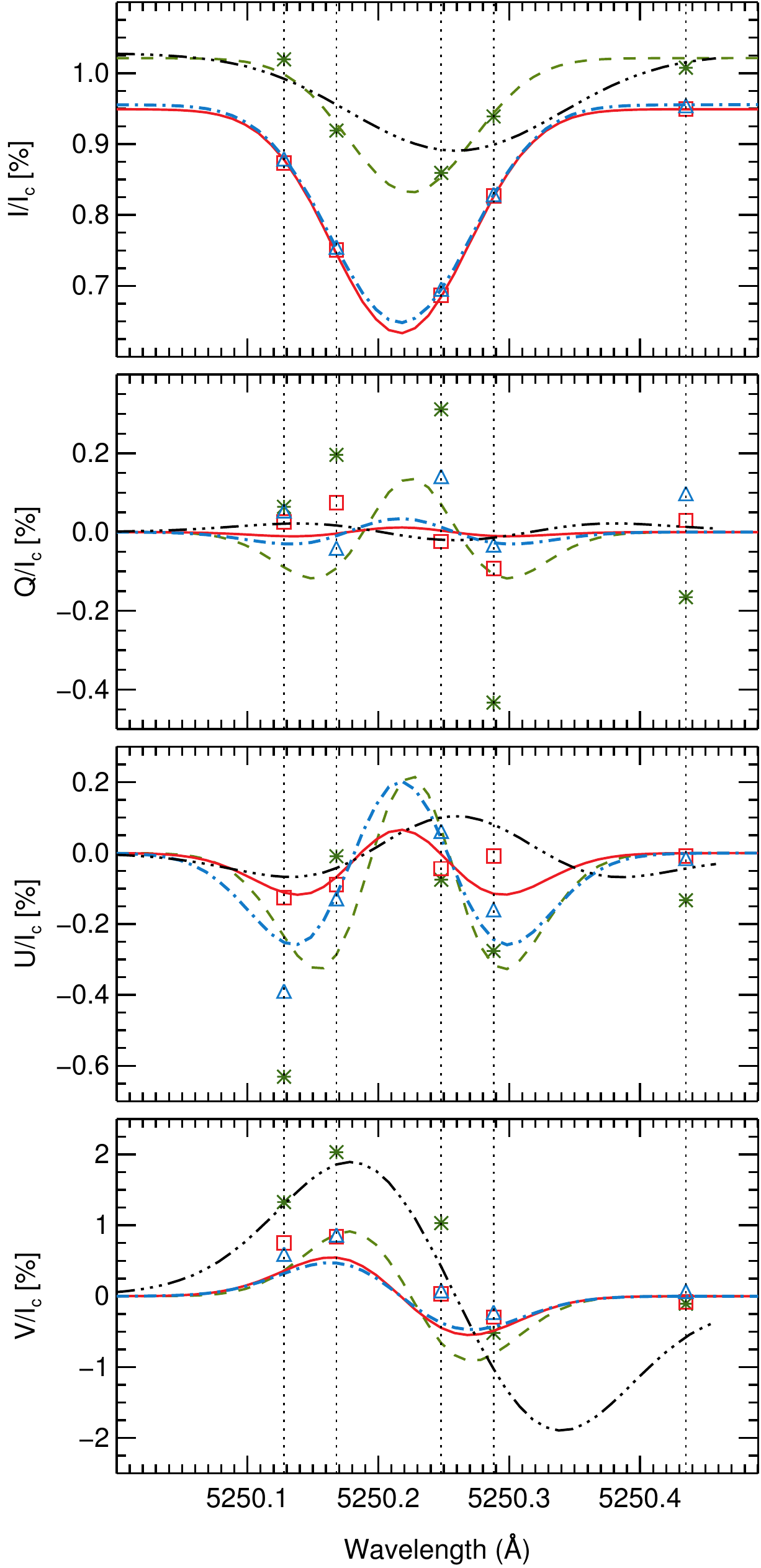}
	\caption{Observed (symbols) and fitted (curves) Stokes $I$, $Q$, $U$, and $V$ profiles for a sample MBP marked in Fig.~\ref{fig:incl-obs}. Red (squares and solid line): non-reconstructed, spatially smoothed IMaX data (see main text); fitted with the SPINOR inversion code. Blue (triangles and dot-dashed line): non-reconstructed data; fitted with the SPINOR code. Green (asterisks and dashed line): phase-diversity reconstructed data; fitted with the SPINOR code. Black (triple-dot-dashed line): fit to the phase-diversity reconstructed data returned by the SIR inversion code. The vertical dotted lines represent the IMaX filter wavelength positions. All profiles are normalised to the IMaX Stokes $I$ continuum.}
	\label{fig:profiles}
\end{figure}

Figure~\ref{fig:subframes} displays maps of different parameters around the sample MBP marked in Fig.~\ref{fig:incl-obs}. The $CP$ and $LP$ maps are based on (non-smoothed) non-reconstructed data, as described in Sect.~\ref{sec-obsincl}. The plotted magnetic field parameters $B_{i}$ and $\gamma_{i}$ were also computed from the non-reconstructed Stokes profiles, using the SPINOR code. The overlaid (red) contours on all panels of Fig.~\ref{fig:subframes} indicate the magnetic patches matching the $4\sigma$ noise level in $CP$ and confirm the magnetic origin of the MBPs observed in the continuum and line-core intensity images. The centre of gravity of the $CP$ signal (located between the formation heights of continuum and line-core) is marked by a cross in each map. We note that the MBP in the continuum image is located on the right-hand side of the cross (with a $0.07~\mathrm{arcsec}$ offset), while in the line-core filtergram it is located on the left-hand side of the cross (with a $0.04~\mathrm{arcsec}$ offset). Thus all images give a consistent picture of an inclined flux tube. This is an example of the test for confirming the determined inclination angle from our geometric approach (see Sect.\ref{subsec-incl-uncertainty}).

It is typical that although the magnetic field patch is bigger and more amorphous than the MBPs, its centre of gravity (marked by the cross) lies close to the centre of gravity of MBPs. The $LP$ signal in this map is almost everywhere below the $1\sigma$ noise level implying that both Stokes $Q$ and $U$ are not significant at the $1\sigma$ level at the position of our MBP. Hence, in principle only an upper limit on the inclination of the magnetic field in this MBP can be given.

The magnetic field strengths and inclination angles exhibit a wide range of values: $18~\mathrm{G}<B_{i}<306~\mathrm{G}$ and $60^{\circ}<\gamma_{i}<87^{\circ}$ among pixels with $CP\geq4\sigma$ noise level, i.e. pixels inside the contours in Fig.~\ref{fig:subframes}. The large $\gamma_{i}$ values may be a result of almost pure noise in the individual Stokes $Q$ and $U$ profiles. We note that the range of inclinations given above is that returned by the inversion code for the MBP under study. In actual fact the inclinations may be considerably smaller.
The $B_{i}$ and $\gamma_{i}$ at the location of the MBP (marked by the cross in Fig.~\ref{fig:subframes}) is found to be $194~\mathrm{G}$ and $75^{\circ}$, respectively. The large $\gamma_{i}$ returned by the inversion at the location of the MBP is incompatible with the small $\gamma_{p}$ (equal to $19^{+11}_{-8}$~degrees) returned by our measurements at the two heights.

In order to investigate the effect of noise level on the computed results, we performed inversions with the SPINOR code on the three sets of differently treated data, described in Sect.~\ref{sec-obsincl}, i.e. PD reconstructed, non-reconstructed, and spatially smoothed (using a boxcar average of $3$ pixels) non-reconstructed data. In addition, to make sure that the large $\gamma_{i}$ values are not an artifact of the SPINOR code, we compared the results of the SIR inversion code performed on the phase-diversity reconstructed data as well as the results of the VFISV inversion code employed on both PD reconstructed and non-reconstructed data.

Figure~\ref{fig:profiles} presents the Stokes $I$, $Q$, $U$, and $V$ spectra recorded by IMaX for the sample MBP marked in Fig.~\ref{fig:incl-obs}. The different symbols represent the differently treated data: green asterisks for the PD reconstructed, blue triangles for the non-reconstructed and red squares for the spatially smoothed non-reconstructed data. The curves represent the corresponding best fits from the SPINOR inversion code: green dashed line for the PD reconstructed, blue dot-dashed line for the non-reconstructed and red solid line for the spatially smoothed non-reconstructed data. The best-fit profiles from the SIR inversion code applied to the phase-diversity reconstructed data are overlaid as black triple-dot-dashed lines. For simplicity, we have not over-plotted the profiles resulted from the VFISV inversion code. However, distributions of the magnetic field parameters that resulted from this inversion code will be later compared with similar distributions computed from the other inversion codes.
Evidently, the fits to the Stokes $Q$ and $U$ signals do not match the observed noisy profiles, which is not surprising given that the linear polarisation signal at the position of this MBP lies below the $1\sigma$ noise level.

Comparing the original with the spatially smoothed non-reconstructed data indicates that the Stokes $V$ signal is hardly affected by the spatial smoothing, while the Stokes $Q$ and $U$ signals are strongly affected by the smoothing of the non-reconstructed data. This is to be expected if the $Q$ and $U$ profiles are dominated by noise.

The magnetic field inclination and field strength values resulting from the inversions whose best-fit profiles were presented in Fig.~\ref{fig:profiles} display a large range of values: $40^{\circ}$-$83^{\circ}$ for $\gamma_{i}$ and $194$-$587~\mathrm{G}$ for $B_{i}$.

\subsection{Statistics and discussion}
\label{statisticsIncl}

\begin{figure}[bht!]
	\centering
	\includegraphics[width=8.7cm]{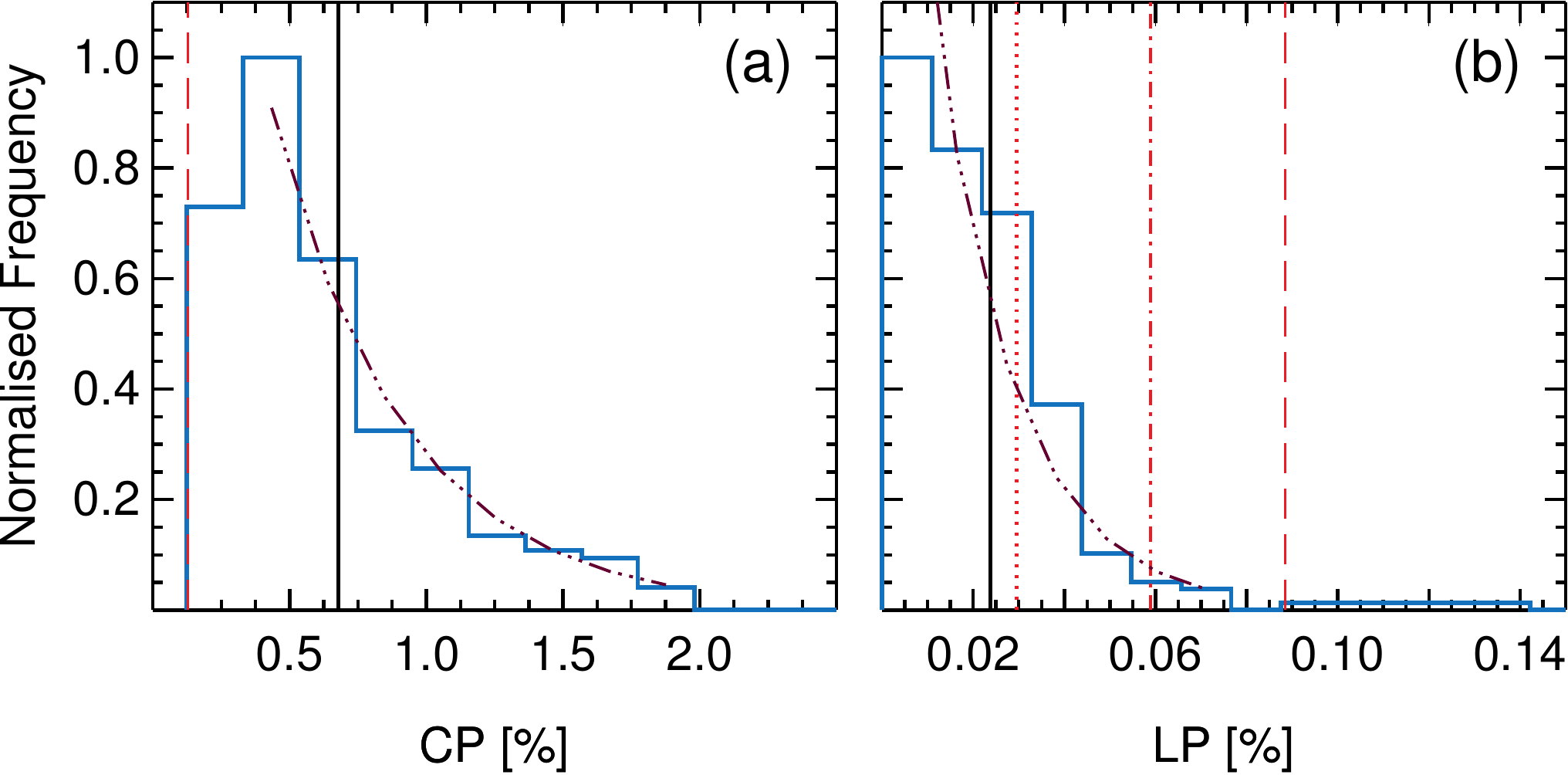}
	\caption{Distributions of the $CP$ (panel $a$) and $LP$ (panel $b$) at the positions of small {\sc Sunrise}/IMaX MBPs. The red dashed line in panel ($a$) marks the $4\sigma$ noise level and the vertical (black) solid lines in both panels represent the mean values of the histograms. The red dotted, dot-dashed and dashed vertical lines in panel ($b$) indicate the $1\sigma$, $2\sigma$, and $3\sigma$ levels, respectively. The triple-dot-dashed curves represent the exponential fits to the histograms (see main text).}
	\label{fig:stat_pol}
\end{figure}

Plotted in Fig.~\ref{fig:stat_pol} are the distributions of the unsigned $CP$ and the $LP$ values (obtained from the least noisy spatially smoothed, non-reconstructed data; described in Sect.~\ref{sec-obsincl}) in all $386$ small MBPs studied here.

The triple-dot-dashed curve in Fig.~\ref{fig:stat_pol}$a$ is an exponential fit to the $CP$ histogram (for $CP>0.45\%$) with an e-folding width of $0.48\%$. The $CP$ histogram shows a lower limit of $0.13\%$ which corresponds to the $4\sigma$ noise level as imposed as one of the selection criteria. In addition to the main distribution with a tail reaching to $1.9\%$, a few larger $CP$ values of up to $4.3\%$ (lying outside the range of this plot) were also obtained. The mean $CP$ value of $0.68\pm0.48\%$ is given by the vertical solid line in Fig.~\ref{fig:stat_pol}$a$. This mean value (obtained from the spatially smoothed non-reconstructed data) is smaller by a factor of $2.8$ than the mean $CP$ measured from the PD reconstructed data, and by a factor of $3.6$ than that of \citet{Riethmuller2013b}, who determined this parameter also from the PD reconstructed data. The small difference in $CP$ values of MBPs to the work of \citet{Riethmuller2013b} is easily explained by the fact that they considered MBPs with a larger range of sizes, whereas we restrict ourselves to the smallest, point-like features.

In contrast to the strong $CP$ signals at the position of our small MBPs, the distribution of $LP$, illustrated in Fig.~\ref{fig:stat_pol}$b$, shows $LP$ signals which almost always lie below $2\sigma$. In particular, $83\%$ of the MBPs have $LP\leq 1\sigma$, $12\%$ have $1\sigma<LP\leq2\sigma$, $3\%$ belong to the range of $2\sigma<LP\leq3\sigma$ and only $2\%$ corresponds to $LP>3\sigma$. The exponential fit, with an e-folding width corresponding to a $LP$ of $0.018\%$, is overlaid as a triple-dot-dashed line (fit limited to $LP \leq 0.08\%$). The largest $LP$ values found in our sample reach up to $0.3\%$ (not shown in the histogram). Therefore, the majority of our selected MBPs have a weak, mostly noise-dominated $LP$ signal with a mean value of $0.024\pm0.018\%$ indicated by a vertical solid line in Fig.~\ref{fig:stat_pol}$b$. The too large fraction of MBPs with $LP<1\sigma$ compared with the expectations for a Gaussian distribution indicates that the deduced noise level is too large. This is to be expected since the noise is an average over the full IMaX field of view, whereas, by definition the MBPs are associated with a continuum enhancement and a weaker spectral line, i.e. with a larger photon flux, so that $\sigma_{noise}$ at the MBPs is expected to be slightly lower.

\begin{figure}[bh!]
	\centering
	\includegraphics[width=8.5cm]{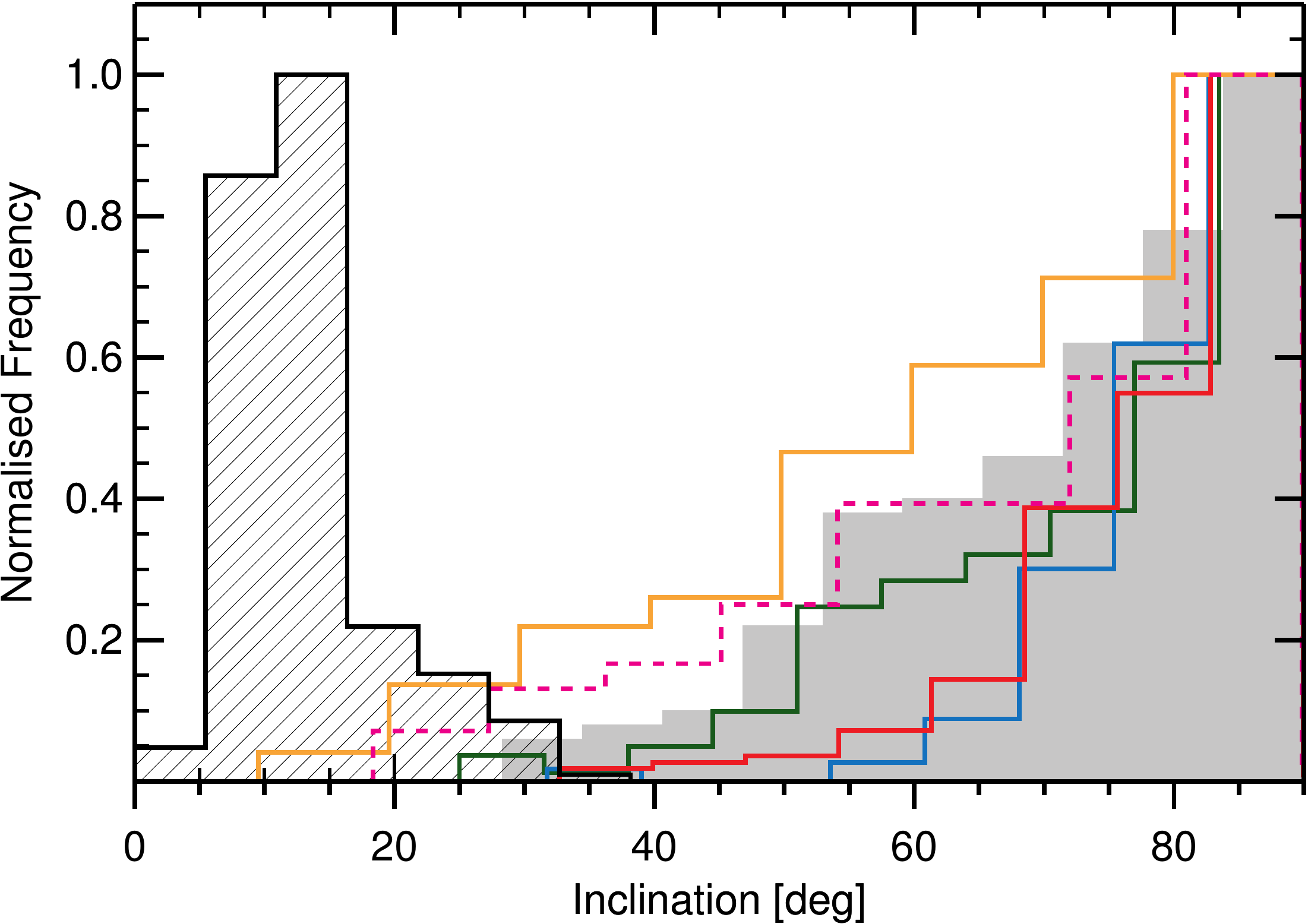}
	\caption{Distributions of inclination relative to the line of sight of MBPs. Inclinations, $\gamma_{p}$, obtained by comparing locations of MBPs in two layers (from the IMaX/continuum-IMaX/line-core combination; see main text) are represented by the (black) hashed histogram. Distributions of magnetic inclination angles, $\gamma_{i}$, of the same magnetic features computed by inverting Stokes data (see main text) are found on the right-hand side of the plot. The different histograms result from inversions employing different codes and applying them to data treated in different ways (see main text for details). All histograms are normalised to their maximum values.}
	\label{fig:incl-stat}
\end{figure}

\begin{figure}[bh!]
	\centering
	\includegraphics[width=8.5cm]{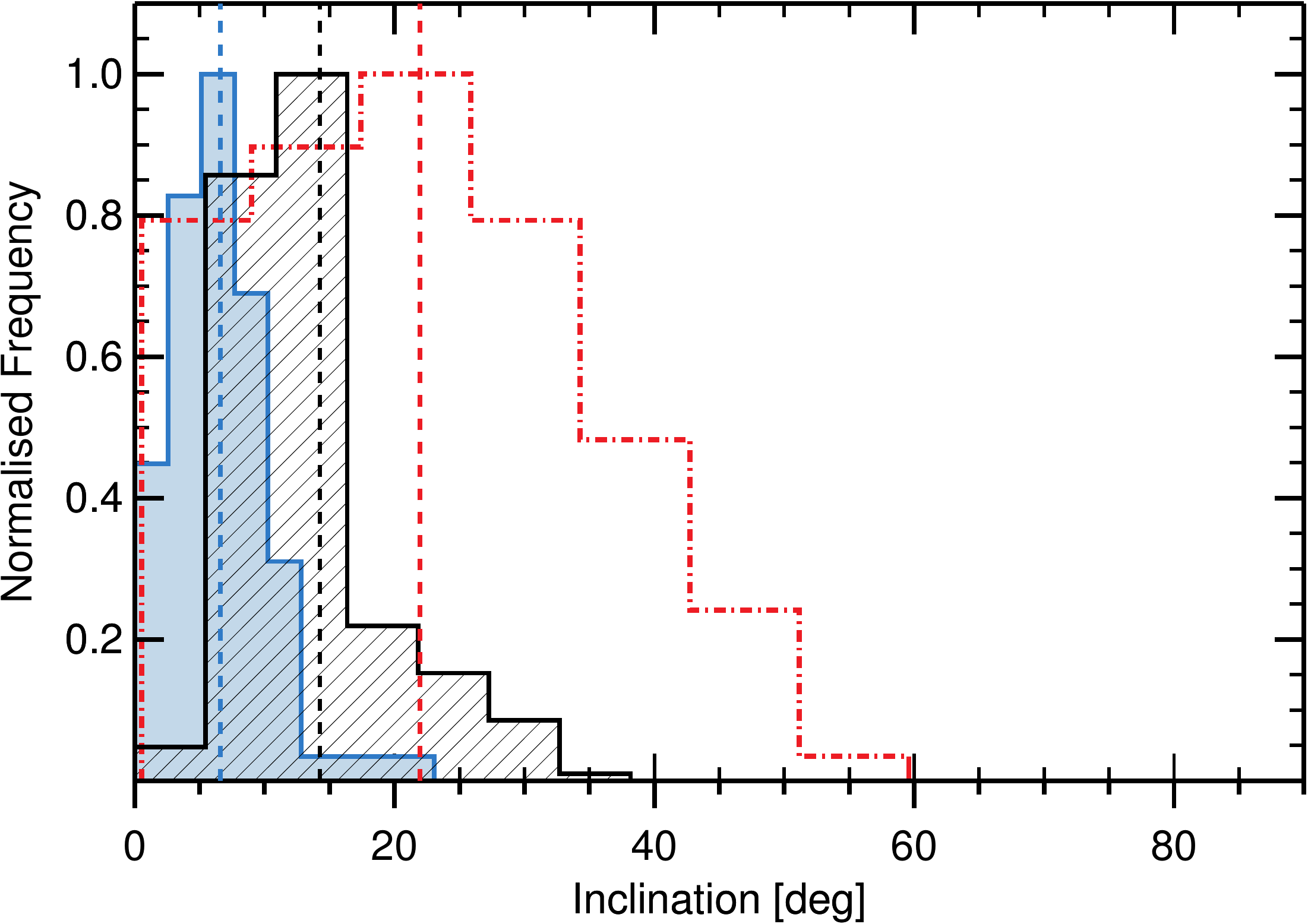}
	\caption{Distributions of inclination of MBPs obtained from the proposed geometric method, $\gamma_{p}$ (see main text). The (black) hashed, (blue) shaded and (red; dot-dashed) outlined histograms represent the distributions of $\gamma_{p}$ obtained from combinations of two atmospheric layers of the IMaX/continuum-IMaX/line-core, the IMaX/continuum-SuFI/Ca~{\sc ii}~H, and the IMaX Stokes $V$ images made in the inner and outer flanks of the line, respectively. All histograms are normalised to their maximum values. The vertical dashed lines indicate mean values of the histograms.
	}
	\label{fig:statgp}
\end{figure}

In Fig.~\ref{fig:incl-stat} the distributions of the magnetic field inclination angle in all $386$ small and isolated internetwork MBPs are plotted, as obtained from our geometric method using the IMaX/continuum-IMaX/line core combination, $\gamma_{p}$ (black hashed histogram on the left side), and from the inversion of Stokes profiles, $\gamma_{i}$ (all the remaining histograms). They reveal a clear discrepancy between the almost vertical fields peaking at $14^{\circ}$ (mean $\gamma_{p}$ of $14\pm6^{\circ}$) obtained from the intensity images and the nearly horizontal magnetic fields (histograms in the right part of the figure), in the same magnetic elements, determined with the inversion codes. The grey shaded and yellow outlined histograms illustrate the distribution of $\gamma_{i}$ obtained from inversions of PD reconstructed data made with the SIR and VFISV codes, respectively. The purple dashed-line identifies the distribution of $\gamma_{i}$ obtained from inverting the non-reconstructed data with the VFISV code. The rest of the outlined histograms on the right-side of Fig.~\ref{fig:incl-stat} represent distributions of $\gamma_{i}$ computed with the SPINOR code from the differently treated data (described as in Sect.~\ref{sec-obsincl}): ($1$) the non-reconstructed images (blue), ($2$) the spatially smoothed non-reconstructed data (red), and ($3$) PD reconstructed images (green). The fact that independent inversions carried out by three codes and applied to differently treated data with different noise levels produce qualitatively similar distributions of $\gamma_{i}$ confirms that the difference between the inclination angles obtained from the geometric technique and those from the inversions is robust.

There are, however, quantitative differences between the various histograms of $\gamma_{i}$. Thus the VFISV code returns somewhat less horizontal field than the other two codes. This indicates that Milne-Eddington inversions may be somewhat less affected by noise than a depth-dependent inversion.

We note that the inversion codes return formal errors, which are generally used as an indication of the uncertainties in the deduced quantities. However, both, the parameters and the formal errors, are not reliable when there is almost no signal in Stokes $Q$ and $U$ (and the Stokes $V$ signal is not very strong). The formal errors returned by inversion codes, that are computed from the uncertainty in fits of the synthesised to observed Stokes profiles, are not reliable in our case, since the $Q$ and $U$ profiles are dominated by noise. Because the code cannot distinguish between the true signal and only noise profiles, it finds the best fit to the noise when there is no significant linear polarisation signal.

When there is no linear polarisation signal, the inversion codes tend to return an inclination tending towards $90^{\circ}$ (horizontal fields) with a small uncertainty. The reason that the inversion returns nearly horizontal fields when there is no linear polarisation signal is the different relationships between the magnetic field and the various polarised Stokes parameters. Thus, in the weak field approximation Stokes $V$ (circular polarisation) is proportional to the field strength while Stokes $Q$ and $U$ (linear polarisations) are proportional to B squared. This means that in the absence of true $Q$ and $U$ signals (i.e. a vertical field), but in the presence of noise (which is interpreted by the inversion as a weak Stokes $Q$ and $U$ signal), the inversion code returns a relatively horizontal field. Thus typical uncertainties returned by the codes are only $2^{\circ}-4^{\circ}$, which is an order of magnitude smaller than the average difference in inclination between our geometric method and the inversions. Thus, this difference is highly significant and suggests that in the absence of $Q$ and $U$ signals under quiet Sun conditions inversion codes return unreliable parameters and unreliable error bars.

In addition, we found (from our geometric method) that the magnetic elements are not preferentially inclined in any particular direction.

Our geometric method reveals the presence of nearly vertical IN magnetic fields, very much in contrast with the rather horizontal fields returned by inversions. The obviously too large inclinations returned by the inversions support the results of \citet{Borrero2011a,Borrero2012}, who showed such a discrepancy between the results of inversions and of both Monte Carlo and numerical simulations on the distribution of magnetic inclination angles when Stokes $Q$ and $U$ are dominated by noise.

For comparison, the distributions of $\gamma_{p}$ obtained using different combinations of the two heights are plotted in Fig.~\ref{fig:statgp}. The black hashed histogram is the same distribution of $\gamma_{p}$ showed on the left side of Fig.~\ref{fig:incl-stat} (from the IMaX/continuum-IMaX/line-core images). The distributions of $\gamma_{p}$ measured from the combinations of IMaX/continuum-SuFI/Ca~{\sc ii}~H and the IMaX Stokes $V$ images made in the inner and outer flanks of the Fe~{\sc i} line are displayed as blue shaded and red (dot-dashed) outlined histograms, respectively. The vertical dashed lines in Fig.~\ref{fig:statgp} indicate the average $\gamma_{p}$ values, lying at $7^{\circ}$, $14^{\circ}$ and $22^{\circ}$ (with standard deviations of $4^{\circ}$, $6^{\circ}$ and $19^{\circ}$ of the various distributions) for the IMaX/continuum-IMaX/line-core, IMaX/continuum-SuFI/Ca~{\sc ii}~H and IMaX Stokes $V$ images, respectively. The three distributions of $\gamma_{p}$ give qualitatively similar results, confirming the generally vertical fields of the MBPs. In Sect.~\ref{subsec-incl-uncertainty} we discussed the possible causes for the quantitative differences among the three distributions of $\gamma_{p}$. In particular, we noted that a smaller height difference between the two layers introduces a larger error/uncertainty in deducing the horizontal position on the derived $\gamma_{p}$.

\begin{figure}[bh!]
	\centering
	\includegraphics[width=8.5cm]{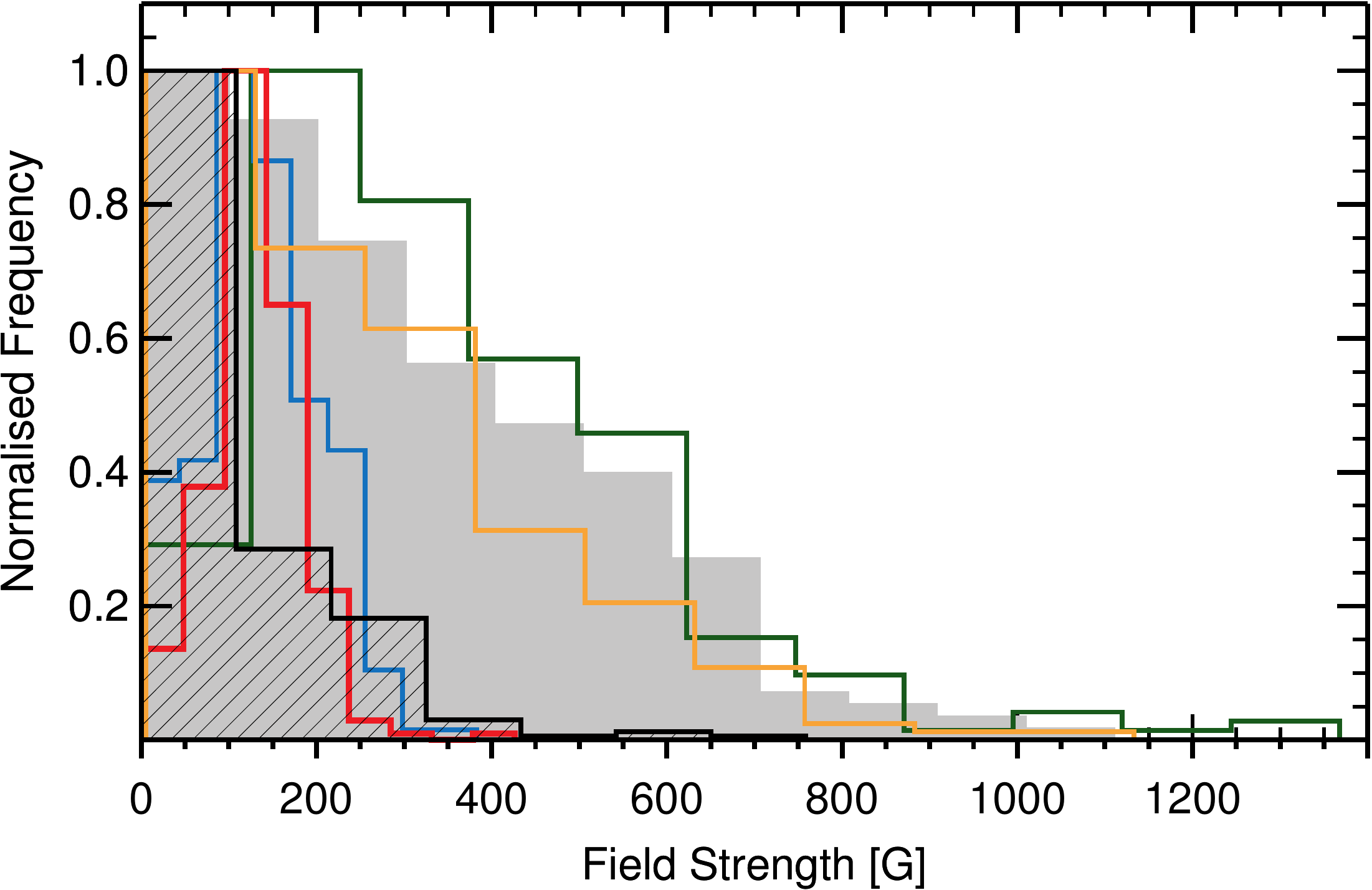}
	\caption{Distributions of the field strength $B_{i}$ deduced in the same MBPs as in Fig.~\ref{fig:incl-stat}. The (black) hashed histogram shows the distribution of $B$ obtained from the SPINOR inversion code with pre-determined inclination angles ($\gamma_{p}$) from measurements at two heights (see main text). Other histograms illustrate the distributions of the field strength of the same magnetic features computed by inverting differently treated data and using different inversion codes (see main text) without imposing $\gamma$. The colours refer to the same inversions as in Fig.~\ref{fig:incl-stat}.}
	\label{fig:stat_b}
\end{figure}

We note that if the inversions derive an incorrect inclination, they possibly also deduce an incorrect pixel-averaged field strength $B_{i}$ (which is at least partially determined by the amplitude of the Stokes profiles, so that an overestimate of the $LP$ responsible for the overestimate of $\gamma$ should also result in an overestimate of $B$). In order to obtain an improved value of $B$, we use the inclination obtained from our geometric method, $\gamma_{p}$, as an input to the inversion of the spectropolarimetric data. The inversion code is forced to ignore fitting the noise-dominated Stokes $Q$ and $U$ and hence, should provide more reliable magnetic field strength for small, IN magnetic elements compared to those obtained from unconstrained inversions.

The distributions of $B_{i}$ computed from spatially smoothed (red histogram) as well as non-smoothed non-reconstructed data (blue histogram) using the SPINOR code, presented in Fig.~\ref{fig:stat_b}, have a mean value of $130~\mathrm{G}$. To avoid cluttering the figure, the distribution of the $B_{i}$ obtained from inverting the non-reconstructed data with the VFISV code (with a mean value of $160~\mathrm{G}$) is not plotted. Inverting the PD reconstructed images results in a mean $B_{i}$ in the range of $260-360~\mathrm{G}$, depending on the inversion code (green, yellow and grey shaded histograms). A mean value of $B_{p}=100~\mathrm{G}$ is obtained from the SPINOR code with pre-determined inclination angles $\gamma_{p}$ from non-reconstructed data (hashed histogram). This is obviously smaller than mean values obtained from unconstrained inversions.

A visual comparison between the distributions of $\gamma_{i}$ obtained from different inversions in Fig.~\ref{fig:incl-stat} shows that the results of the SPINOR inversion from the non-reconstructed data (i.e. from both non-smoothed and spatially smoothed data; blue and red histograms) represent larger $\gamma$ compared to other distributions. Consequently, larger $B_{i}$ values would be expected for the blue and red histograms in Fig.~\ref{fig:stat_b} compared to the others. However, their distributions show smaller $B_{i}$ values. This non-compatibility arises because the magnetic field is more diffuse and is spread over a larger area in the non-reconstructed data. Hence, $B_{i}$ in a single pixel (selected at the position of the MBPs in non-reconstructed data) is expected to be smaller than the value in PD reconstructed data. The effect of the different spatial resolution more than offsets influences the difference in $\gamma_{i}$.

We now estimate by how much the contribution to the horizontal magnetic flux due to these MBPs is overestimated by inversions. Horizontal magnetic flux ($\Phi_{h}$) at any given pixel can be obtained using the computed $B$ and $\gamma$ values, i.e. $\Phi_{h} \propto B sin(\gamma)$. Hence, larger $B$ and/or larger $\gamma$ result in larger $\Phi_{h}$. The ratio between the horizontal magnetic flux computed from the inversion results ($\Phi_{h,i}$) and the one obtained from our geometric method ($\Phi_{h,p}$) can be approximated as
\begin{equation}
\frac{\Phi _{h,i}}{\Phi _{h,p}} = \frac{B_{i}\: sin(\gamma _{i})}{B_{p}\: sin(\gamma _{p})}\,,
	\label{equ:fluxh}
\end{equation}
\noindent
where $B_{i}$ and $\gamma_{i}$ are the estimated field strength and inclination angle from the inversions, respectively. $\gamma_{p}=14^{\circ}$ is the mean inclination from the geometric technique and $B_{p}=100~\mathrm{G}$ represents the field strength computed using the inversion with pre-determined $\gamma_{p}$. As a result, the mean values of the $B_{i}$ and $\gamma_{i}$ from different inversions reveal that the horizontal magnetic flux is on average overestimated by a factor of $5-15$ compared to $\Phi_{h,p}$.

However, we note that none of these magnetic elements could be observed as MBPs unless they would be $\mathrm{kG}$ magnetic concentrations~\citep[e.g.][]{Rutten2001,Voegler2005,Riethmuller2013a,Riethmuller2013b}. This means that these MBPs are likely not fully resolved and hence, all the field strength values that we computed using inversions were underestimated. Therefore, a much smaller filling factor than unity would be needed to obtain the true $B$ values. This reasoning has no effect on the ratio of fluxes deduced in the previous paragraph.

\section{Conclusions}

We propose a simple technique for determining the inclination of magnetic elements associated with bright points by comparing the locations of the MBPs in high spatial resolution intensity images sampling two heights, or, alternatively by comparing the locations of peak Stokes $V$ signal at two heights. This technique is applied to {\sc Sunrise}/IMaX and SuFI data. The method offers, for the first time, an opportunity to determine the inclination angle in small-scale magnetic features independently of inverting measured Stokes profiles. The new technique is of particular interest in the quiet-Sun since the Stokes profiles can be affected by noise there, making inversions less reliable.

For a first application of our technique we selected small, point-like bright features (diameter smaller than $0.3~\mathrm{arcsec}$) displaying a $CP$ (a measure of Stokes $V$ strength) above the $4\sigma$ noise level. They turned out to have very weak linear polarisation signals, with $LP$, computed from Stokes $Q$ and $U$ profiles, lying almost always below the $2\sigma$ level.

The high spatial resolution and seeing-free data recorded by {\sc Sunrise} allowed the accurate position of a MBP associated with one and the same magnetic element to be determined in at least two different layers. We have employed the continuum position and the line-core of the {\sc Sunrise}/IMaX Fe~{\sc i} $5250.2~\mathrm{\AA} $ passband, but have also compared {\sc Sunrise}/SuFI Ca~{\sc ii}~H brightenings with those in the $5250.4~\mathrm{\AA} $ continuum. The inclination is obtained by connecting the coordinates of a pair of MBPs associated with the same magnetic feature (i.e. the same $CP$ patch) and considering the formation heights of the passbands in which the MBPs are observed. There are a number of sources of uncertainty, such as the time difference between data recorded at different wavelengths and the fact that the formation heights depend on the atmospheric structure (e.g. temperature, electron pressure and magnetic field). However, using reasonable estimates of these and other uncertainties results in inclination angles accurate to better than $11^{\circ}$.

An application of this method to $386$ small magnetic features in the IN quiet-Sun gives an average inclination of $14^{\circ}$ and a standard deviation of only $6^{\circ}$. Employing Ca~{\sc ii}~H line-core images (less sensitive to Zeeman effect and Doppler shifts) instead of the Fe~{\sc i} $5250.2~\mathrm{\AA} $ line-core gave very similar results, providing even somewhat more vertical fields ($\gamma_{p} \approx 7^{\circ}$ on average). Our results based on intensity images were generally confirmed by comparing spatial centre-of-gravity of the Stokes $V$ signal at two different positions in the line, formed at somewhat different heights, giving $\gamma_{p} \approx 22^{\circ}$ on average. This last test is of particular important in spite of the larger errors it gives (because of the lower signals, spatially less compact structures in Stokes $V$ and the much smaller height difference), since it reveals that the geometric method applied to MBPs does provide a good estimate of the inclination of the magnetic field.

There is very little overlap between the distributions of inclination obtained with our geometric technique and from the three Stokes inversion codes, which gave average inclinations $\gamma_{i}$ of $66^{\circ}-81^{\circ}$, with all $\gamma_{i}$ distributions peaking at or close to $90^{\circ}$. The striking agreement between the various inversions supports the suggestion that inversion codes overestimate the inclination angles of features with noise-dominated Stokes $Q$ and $U$ signals~\citep[e.g.][]{deWijn2009,Borrero2011a,Borrero2012}. Inversions of differently treated data (i.e. phase-diversity reconstructed, non-reconstructed and spatially smoothed non-reconstructed data) and three independent inversion codes gave similar results.

The results of such measurements at the two heights are also found to have a significant effect on determining the solar magnetic flux in horizontal fields, due to the studied small magnetic elements. Our work indicates that traditional inversion methods overestimate this parameter by an order of magnitude, at least for the field associated with small MBPs.

Furthermore, we found that the magnetic field strength (on the Sun) computed from the inversions of small, IN magnetic elements is overestimated. We showed that the inversions give a lower field strength by an average factor of $\approx2$ when the inclination angle, $\gamma_{p}$, obtained from our geometric method is imposed prior to inverting the data.

We have restricted ourselves to small magnetic elements that manifest themselves as BPs, which are expected to be $\mathrm{kG}$ magnetic concentrations~\citep[e.g.][]{Spruit1976}. Only a concentrated field produces a sufficiently deep Wilson depression to allow enough excess radiation to enter the magnetic feature to produce a continuum BP. Thus, \citet{Riethmuller2013b} use MHD simulations to demonstrate that only magnetic features with $\mathrm{kG}$ field produce significant continuum brightenings. The large expected intrinsic field strength is consistent with the small $\gamma_{p}$ found here, since $\mathrm{kG}$ fields are expected to be relatively vertical due to their buoyancy~\citep{Schuessler1986}. The comparison of the deduced field strengths of roughly $100~\mathrm{G}$ with the requirement of $B>1000~\mathrm{G}$ means that the diameter of the magnetic features studied here is on average smaller than $30-40~\mathrm{km}$ assuming a resolution element of {\sc Sunrise}/IMaX of $100~\mathrm{km}$ (after reconstruction). Thus, although the high spatial resolution observations from {\sc Sunrise}/IMaX have allowed us to resolve magnetic elements in the quiet-Sun~\citep{Lagg2010}, many small-scale magnetic elements observed by {\sc Sunrise} are likely to be spatially unresolved~\citep{Jafarzadeh2013a,Riethmuller2013b}.

It has recently been shown that only $1/4$ of the IN areas have strong linear polarisation signals, i.e. signals above the $4.5\sigma$ noise level~\citep{Orozco2012b}. Therefore, the inclination of the magnetic vector in, at least, $3/4$ of the IN area (i.e. the majority of the solar surface) is still not clear, with arguments being made for a mostly horizontal field \citep{BellotRubio2012} as well as an isotropic distribution of weak fields~\citep{Asensio2009,Bommier2009}, or even predominantly vertical fields in the quiet-Sun~\citep{Stenflo2010,Stenflo2013}.

Our results demonstrate that at least some of the magnetic features indicated by inversions to harbour nearly horizontal fields are actually close to vertical (see Fig.~\ref{fig:incl-stat}). This requires a reassessment of the distribution of the magnetic field vector, especially in regions where Stokes $Q$ and $U$ are highly affected by noise. The method proposed here can help by providing inclinations for all magnetic features associated with MBPs (i.e. strong-field elements). In the absence of significant Stokes $Q$ or $U$ signals, the inversion can be constrained by using the inclination angles deduced from the geometric method, in order to obtain better values of the field strength. The combination of both methods then allows the full vector magnetic field to be inferred (the strength from the inversion, the inclination and azimuth from the geometric method). This would be impossible with inversions alone in the case of MBPs.

We can foresee a wide applicability of the new technique. Applying this method to higher contrast images, e.g. photospheric $2140~\mathrm{\AA} $ and Ca~{\sc ii}~H $3968~\mathrm{\AA} $ obtained by {\sc Sunrise}/SuFI~\citep{Riethmuller2010,Gandorfer2011}, and loosening constraints on the size of the considered MBPs will result in better statistics. An even wider applicability of the new technique would result from measurements of the location of peaks (or centre-of-gravity) in the Stokes profiles (generally Stokes $V$ because of its larger S/N value) at measurements made at different heights. We tested this by measuring the location of magnetic concentrations (centre-of-gravity) observed in the Stokes $V$ images in two wavelength positions of the Fe~{\sc i}~$5250.2~\mathrm{\AA}$ line formed at somewhat different heights and found that the results are consistent with those obtained from the intensity images at two atmospheric layers. We also note that the Fe~{\sc i}~$5250.2~\mathrm{\AA}$ line may not be so ideal for this approach due its very significant thermal weakening in magnetic elements. This causes larger uncertainties in estimating height differences and hence in measuring inclination angles. Measurements in multiple spectral lines with sufficiently different heights of formation can increase the reliability of the method and may one day even allow the curvature of the magnetic elements (i.e. the bending of the axis of the underlying flux tube with height) to be determined.

\begin{acknowledgements}

The German contribution to {\sc Sunrise} is funded by the Bundesministerium f\"{u}r Wirtschaft und Technologie through the Deutsches Zentrum f\"{u}r Luft- und Raumfahrt e.V. (DLR), Grant No. 50 OU 0401, and by the Innovationsfond of the President of the Max Planck Society (MPG). The Spanish contribution has been funded by the Spanish MICINN under projects ESP2006-13030-C06 and AYA2009-14105-C06 (including European FEDER funds). The HAO contribution was partly funded through NASA grant NNX08AH38G.

\end{acknowledgements}

\bibliographystyle{aa} 
\bibliography{ref} 

\end{document}